# Game theoretic pricing policy in online media delivery platforms

Anubhab Banerjee[1] and Carmen Mas Machuca[2]

[2]Department of Electrical and Computer Engineering, Technical University of Munich
[1]Department of Informatics, Technical University of Munich
[1]anubhab.banerjee@tum.de

***Abstract*—Residential users get most of their preferred content (e.g., films, news) from their wireless access point, which is connected to content providers through one or several network providers. Nowadays, these access points have enough storage and wireless transmission capacities to distribute the content to neighboring access points if required and allowed by the user. Based on this concept, this paper considers a *hybrid* content distribution model in peer-to-peer content deliver networks. A user or peer in this network can get its content either directly from the content provider (CP), or from a neighboring user who has this content stored and is willing to share it. Two types of users are considered: *premium users* are the users getting always the content directly from the CP, and *standard user* are the users getting the content from other *standard* users, when possible, otherwise from the CP. Furthermore, *standard user* share their content to other users if required. This paper studies the pricing problem of the CP such that there are enough *standard* users to distribute the content (using less network resources and reducing significantly the delay) while maximizing the CP benefits. We address the problem from a game theoretic approach and evaluate when the Nash Equilibrium can be attained. The obtained case have been applied to a real life study using a BBC iPlayer dataset and an analysis on the CP benefits dependence on different parameters. Finally we extend this pricing model when considering also the reduction of load on its content custodian (server) and discuss how the pricing policy is affected.**

## I. Introduction

Throughout the last decades, the media industry is shifting from delivering content via physical medium to digital medium. After the commercialization of Internet in 1994, there has been an exponential growth in digital media content delivery services [1]. According to a CISCO report [2], almost 84% of the internet traffic will be video content by the end of this year (2019).

In the media distribution industry, the three most important players are:

1) Content provider (CP), such as Netflix, is the owner of the content and takes care of its distribution. CP has full control over the content and is solely responsible to determine all the attributes (such as pricing and distribution policy) related to the content.
2) Network provider (NP), such as Deutsche Telekom, owns the network that delivers the content from the CP servers to the users.
3) Content consumer, i.e., users, download and use (e.g., listens, watches) the content. Users pay for the content and acquire it from CP through a NP.

Primarily, there are three types of content distribution models: centralized, decentralized and hybrid. The centralized model is implemented as a client-server system where the content placement, content pricing and customer relationships are centrally managed by the CP without being biased by users. CP can set or change the price of the content and distribution policy according to its own strategy. To get the content, a user has to buy it with a price set by CP. The disadvantage of this model is that, as the content is available only in CP's server, the load on CP's server remains very high all the time if it is a popular content. Also, when the server is multiple hops away from the user, the latency for content delivery increases.

The second model is the decentralized model, which is implemented as a peer-to-peer (P2P) network where the users participate into the process of content distribution by trading content [3]. In this case, a user can store the downloaded content in any cache-enabled home device (e.g., Apple TV) and share it with other users who are interested in the same content. The advantage of this model is that it reduces the load of the CP's servers up to a certain extent [4]. However, the disadvantage from the CP perspective is that the CP is not the sole owner of the content. Thus, while determining the price of the content, the CP has to take user's preferences into account.

A hybrid content distribution model is a combination of both centralized and decentralized models [5]. In this model, a user can choose whether to get the desired content in the centralized or the decentralized way, that is, directly from CP or from other users. The price of the content is different in these two cases. From the perspective of CP, this model finds a compromise between centralized and decentralized model: it reduces the load on CP's server up to some extent, but at the same time, CP looses control on the content pricing and distribution. From user perspective, this model offers user the choice on where to get the content from and how much price user pays for it.

In this paper, for the sake of clarity but without losing generality, we refer to a hybrid content distribution model based on the Wi-Fi enabled content sharing architecture [6], [7]. In this architecture, users are connected to the CP through a fixed network provided by NP and also to other used through Wi-Fi technology. In this hybrid model, CP offers the users two options: *Premium* and *Standard*, which correspond to the centralized and distributed content distribution models respectively. If a user selects the *Premium* option, he/she can directly download the content from the CP's server and he/she does not have to store it nor share it with other users. In this case, the CP has the responsibility of delivering the content, which involves various costs for CP, as for example, the fee to the NP. On the other hand, if a user selects the *Standard* option, he/she must check the availability of the content with other *Standard* users before requesting CP for it. If the content is available with any other *Standard* user, the content delivery takes place from one user to another user and does not incur any cost to the CP. It can be safely assumed that a user selecting the *Standard* option always agrees to share any stored content and has all necessary devices and infrastructure.

**Motivation:** Based on our previous works [6], [7] on a content placement strategy from the perspective of CP in a hybrid content distribution model, we provide in this paper a content pricing policy from the perspective of CP to achieve a viable and advantageous pricing for both CP and users.

In the hybrid content distribution model, from the user perspective, selecting the *Standard* option is the cheapest choice but may not be the best one. For example, if we consider a hybrid video sharing architecture [6], [7], a user may not be able to watch HD videos when selecting the *Standard* option as it depends on the link capacity and availability between users. The same user will be able to do so when selecting the *Premium* option. So the choice of selecting *Premium* or *Standard* option entirely depends on the user's willingness to pay for the content and the content itself. So, it is difficult for CP to predict accurately which option a user will choose and hence, which pricing policy the user will consider.

Now, both *Premium* and *Standard* options have advantage and disadvantage from the perspective of CP. If more users select the *Premium* option, the revenue of the CP increases but at the same time, the load on the CP's server also increases, which causes an increase of the maintenance cost and energy consumption. On the other hand, if more users select the *Standard* option, the load on the CP's server reduces, but also the revenue generated by CP decreases. Our main target is to propose a pricing policy from the perspective of a CP, which is robust against the uncertainty of the number of users choosing the *Premium* and *Standard* options. Although this pricing policy has been developed having the Wi-Fi enabled content sharing architecture [6], [7] in mind, it is to be noted that it is valid for any kind of hybrid content distribution network.

**Our contribution:** The main contributions of this paper are:

- Find the price of the content when the benefits of both, CP and users, are maximized and the equilibrium is obtained when possible.
- Identify the conflicts of interest between CPs and users, and determine the cases where an agreement between them is feasible. In those cases, the price is determined by the Nash Bargaining Solution (NBS).
- Apply the proposed solution in a real case study (using real life data in UK) to observe the impact of the different parameters in the pricing policy.
- Extend the pricing model to include the load reduction on CP's server.

To the best of our knowledge, this work is the first one to introduce an NBS based pricing policy for a hybrid digital content distribution system.

The paper has been organized as follows: we discuss prior research works related to digital content distribution in Section 2. In Section 3 we elaborate the problem, we study in the paper, with example. In Section 4 we discuss three different cases and propose a pricing policy in each case. We also incorporate a study based on real life data for analyzing the pricing policy in Section 5. As an extension of the discussed pricing model, we also include another important parameter (load reduction on CP's server) and update the pricing policy in Section 6 with concluding the paper in Section 7.

## II. Related Work

There have already been previous research works on the pricing of digital content in centralized ( [8]–[12]) and decentralized distribution models ( [13]–[15]), but there is no prior research work on the hybrid distribution model addressed in this paper.

Several works have addressed digital content pricing in centralized distribution model. For example, in [8] the authors explore about Music as a Service and find out that next to price, contract duration and product qualities are two most important product attributes. Their research also studies configurations on consumers' utility and their willingness to pay for premium offers. The paper [9] compares between two business models. In the first, free of access is provided with low product quality and advertisements, while in the second the access is charged and product is of high quality. It has been shown that if the users are advertisement tolerant, the first policy is more profitable than the second one. Researchers in [10] studied the same thing and showed that companies are better off providing both pricing and advertising options to consumers. They suggested that as advertisement revenue rate increases, advertising level should be kept low and based on the factors that affect the price and advertising decisions, they derived the optimal price and advertising level. In [11], two kind of policies - fee based and free with advertisement, were studied in detail and was shown that the mix group that uses both free and fee strategies are the most profitable ones. In [12] the authors analyzed the pricing schemes and DRM (digital right management) protection policy with respect to different collaborative market structures. As the piracy is closely associated with the objects and the distribution channel, they also examine the impact of content quality and network effects on the development of strategies based on all these factors. But this is a model for host centric network only, no discussion for peer-to-peer networks.

Significant research has also been conducted on digital content pricing in decentralized model also. In [13] the researchers proposed a game theory based pricing policy for job allocation in mobile grid networks. They introduced a game theoretic framework to obtain pricing model between a Wireless Access Point server and a mobile device. Along with that, they also introduced a game theory based workload allocation scheme for the mobile devices to maximize the revenue. In [14] the authors proposed a lottery based P2P pricing scheme. This model provides higher revenue for peers who intend to participate and contribute with other peers and higher cost for those who choose to behave selfishly. But the service provider is not included in the game, so the pricing policy of the content provider is not taken into account. The paper [15] compares between two models of digital content distribution - client-server model and peer-to-peer model. They described a monopolistic pricing scheme for distributing digital content over peer-to-peer networks that rewards peer users who participate in the distribution process and showed that peer-to-peer model is more profitable if pricing mechanism provides strong incentives to users to share content.

All the previous research works described above are based on the pricing and distribution of digital content in centralized and decentralized content distribution model. To the best of our knowledge, this work is the first one providing a game theoretic pricing policy in a hybrid content distribution model, i.e., where a user can choose either the centralized or the decentralized model to get the content.

## III. Problem Statement

In this section, the pricing problem considered in this paper is formulated. At first we discuss the background necessary for formulation of this problem. Then objective functions are defined, constraints are discussed and a detailed example is given for better understanding.

### A. Background

As stated earlier, the content distribution network usually involves three players to deliver the content, as shown in Fig. 1: content provider (CP), network provider (NP) and the user (referred also as content customer (CC)). Let us consider a content distribution model of a single CP, a single or multiple NPs and multiple users. All users are connected to the CP via fixed network(s) provided by NP(s). If there are multiple NPs in the model, we assume that each NP charges the same amount of money ($p_n$) to CP. In addition to that, every user is connected to his/her neighboring user(s) via Wi-Fi and is reachable by every other user in single or multiple hops.

Let us denote $x$ to be the total number of users who get a particular content $c$ from CP. These users can select one of the following two options:

- *Premium* service: This option guarantees the user to get the content from the server of CP at a cost of $p_b$ [€ /Byte]. This cost includes the cost of the original content as well as the energy cost associated to the content delivery. The users, who select this service, are called *Premium* users (shown in blue in Fig. 2).
- *Standard* service: This option allows the user to get the content from other *Standard* neighboring users, who have the appropriate content. Any *Standard* user agrees on sharing its content with any other interested *Standard* user. The cost that a *Standard* user has to pay to a content $c$ is $p'_b < p_b$ [€ /Byte]. However, this cost does not include the energy cost, which should be charged to a user only in case of content sharing. The consumed energy is proportional with the size of the content. Let us denote energy cost per transmitted Byte with $s$ [€ /Byte] which is charged to the *Standard* user who is sharing. Furthermore, in order to encourage the content sharing among users who select the *Standard* option, CP not only charges less for getting the same content ($p'_b < p_b$) but also offers a compensation price $p_u$ [€ /Byte] every time that a user shares the content with another user. The users, who select this service, are called *Standard* users (shown in green in Fig. 2).

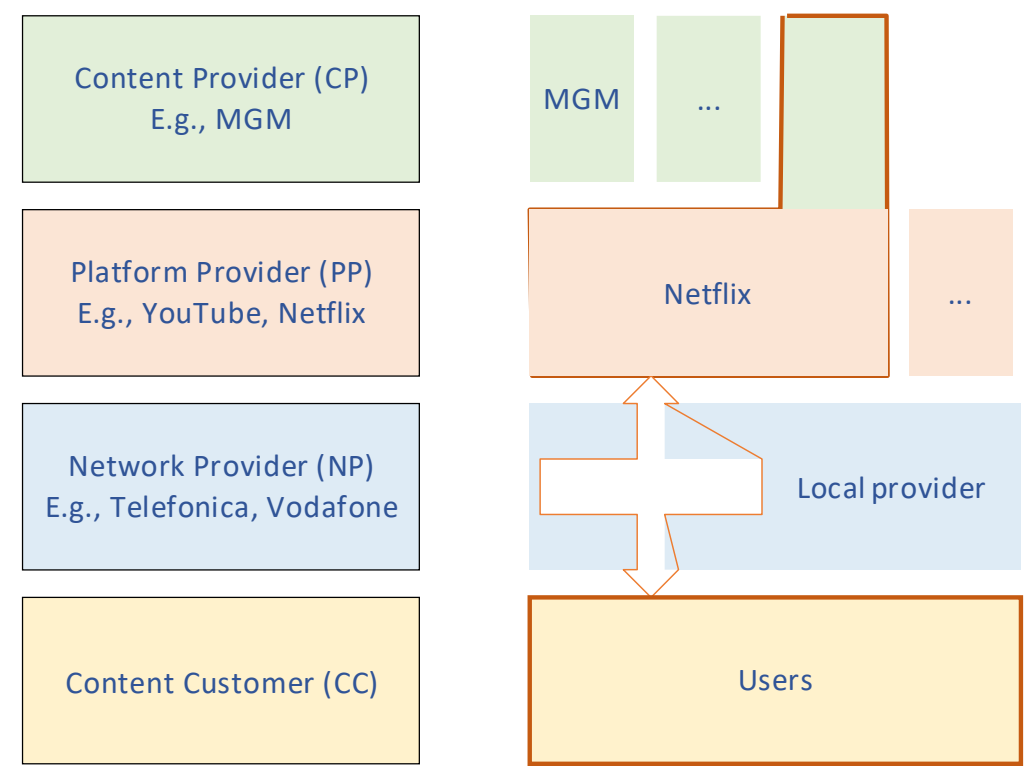

Fig. 1: Top-down Content distribution: from content provider to user, with highlighted players of the game.

Let us consider, that $y$ out of $x$ users are *Premium* users and hence, $x - y$ are *Standard* users.

### B. Objective Functions

In this paper, we formulate a bargaining framework only among those who already own $c$ and who can share it with others. The users, who do not own the content, have no leverage for bargaining and they are kept outside of the bargaining framework. Similarly, as a *Premium* user does not share content, no *Premium* user can take part in the bargaining. As it is a connected network, we also assumed that the *Standard* user who gets the content first is the only one to share it with other interested *Standard* users. The problem that we want to address is to find the content costs $p_b$, $p'_b$ and sharing discount $p_u$ such that benefits for the user ($Ben_{user}$) and for the CP ($Ben_{CP}$) are maximised. Benefits are defined as the difference between revenues and costs. For a given content $c$, both $Ben_{CP}(\cdot)$ and $Ben_{user}(\cdot)$ can be given by the following

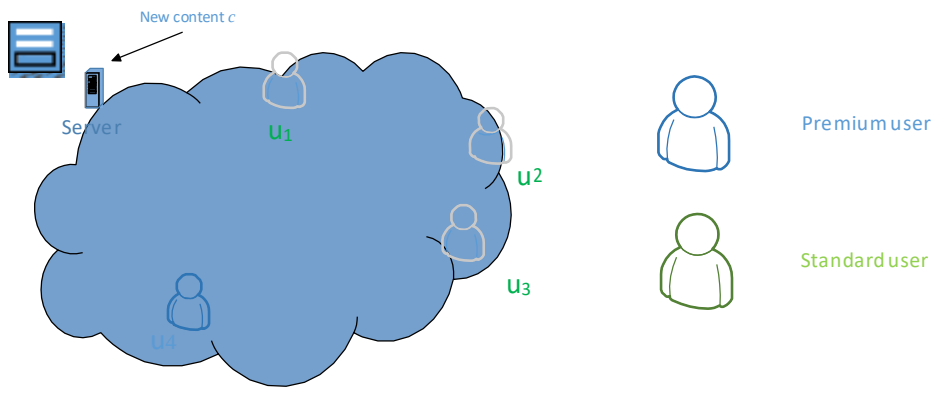

Fig. 2: Case example with $x = 4$ users: $y = 1$ and $x - y = 3$ users getting *Premium* and *Standard* access respectively.

| Notation | Description |
|---|---|
| $f$ [Bytes] | File size |
| $x$ | Total number of users who download the file |
| $y$ | Number of users who download the file via premium option |
| $k$ | Percentage of users who download the file via premium option |
| $p_n$ [e/Byte] | Price paid by CP to network provider to deliver a content from server to client |
| $p_b$ [e/Byte] | Price a user pays to CP when the user gets the content selecting premium option |
| $p'_b$ [e/Byte] | Price a user pays to CP when the user gets the content selecting standard option |
| $r$ | Ratio between premium and standard service cost |
| $p_u$ [e/Byte] | Reward given to user by CP who shares the content in standard option |
| $s$ [e/Byte] | Energy cost charged to the *Standard* user per Byte when sharing the content with other *Standard* users |

TABLE I: List of variables used in the content distribution model

equations (derivation of these equations are given in the next Section):

$$Ben_{CP} = \sum_{j=1}^{x} f[(p_b - p_n)\,a_j + (p'_b - (y_j - 1)\,p_u)\,(1 - a_j)]$$

$$Ben_{user_j} = f[(-p_b a_j - p'_b\,(1 - a_j)) + \beta_j\,(1 - a_j)\,(y_j - 1)\,(p_u - s)]$$

where

$$a_j = \begin{cases} 1, & \text{user } j \text{ gets content as Premium user} \\ 0, & \text{otherwise} \end{cases}$$

and

$$\beta_j = \begin{cases} 1, & \text{Standard user } j \text{ is the first one to get the content} \\ 0, & \text{otherwise} \end{cases}$$

and $f$ is the size of content $c$. Throughout this paper we consider a particular content $c$ and the parameters associated with $c$ are summarized in Table I. The problem should be solved for any content offered to the customers and hence, the price will depend on the number of premium and standard users for each content.

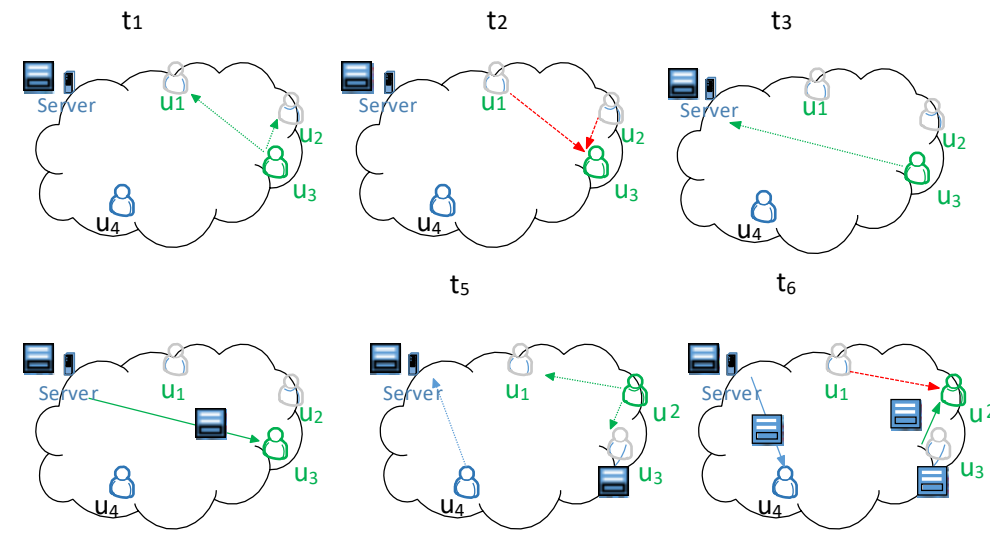

Fig. 3: Steps of content distribution in the hybrid distribution model

## C. Constraints

In this problem, the values of $x$, $y$, $f$, $p_n$, $s$ for a particular $c$ are constant. The pricing model is limited by the following constraints:

- $p_b > p'_b$: This constraint due to the fact that *Premium* users get content directly from CP's server and it incurs additional cost to CP compared with the *Standard* users.
- $p'_b > p_n$: Let us consider the worst case scenario, when only one user watches the content, i.e., $x = 1$. Hence, this user cannot share the content with any other user. In this case, the total benefit of CP $Ben_{CP}$ will be limited to $Ben_{CP} = f.(p'_b - p_n)$. As the CP would always aim at generating positive benefit, the following constraint should hold

$$p'_b > p_n. \tag{1}$$

## D. Example

Let us describe the proposed hybrid content distribution model with an example. Let us consider a new content $c$ of size $f$ Bytes that has been introduced by the CP at a certain point of time $t_0$. We also consider a group of $x = 4$ users $u = \{u_1, u_2, u_3, u_4\}$ who are interested in $c$ and among them, user $u_4$ is the only *Premium* user (i.e., $y = 1$). Let us present the exchange of messages when the first user ($u_3$) requires content $c$.

**$t_1$:** At this time instant, $u_3$ decides to download $c$. As $u_3$ is a *Standard* user, he/she has to check the availability of $c$ with other *Standard* users (in this case, $u_1$ and $u_2$). So, $u_3$ sends request to both $u_1$ and $u_2$ to check if $c$ is available at any of them.

**$t_2$:** As $c$ is not available with $u_1$ or $u_2$, both of them send negative response to $u_3$.

**$t_3$:** $u_3$ then sends a request to CP to download $c$ directly from CP's server as $c$ is not available with any other *Standard* user in the network.

**$t_4$:** $c$ is delivered to $u_3$ via fixed network provided by NP. $u_3$ pays CP an amount of $f.p'_b$, while NP charges CP an amount of $f.p_n$ to deliver the content to $u_3$.

**$t_5$:** Let us consider that at this time, two users ($u_2$ and $u_4$) decide to get $c$. However, $u_2$ decides to get it as a *Standard* user and $u_4$ as a *Premium* user. As $u_4$ is a *Premium* user, $u_4$ directly sends a request to CP to download $c$ from CP's server. However, as $u_2$ is a *Standard* user, first $u_2$ checks the availability of $c$ with other *Standard* users (in this case, $u_1$ and $u_3$).

**$t_6$:** $c$ is delivered to $u_4$ via fixed network provided by NP. $u_4$ pays CP an amount of $f.p_b$. NP charges CP an amount of $f.p_n$ to deliver the content to $u_4$. On the other hand, $u_2$ gets $c$ from $u_3$ via Wi-Fi sharing and pays CP an amount of $f.p'_b$. This sharing cost $u_3$ an amount of $s.f$ for energy consumption. At the same time, $u_3$ gets a reward of $f.p_u$ from CP for sharing the content with $u_2$.

If, at a later point of time $t_7 > t_6$, $u_1$ wants to download $c$, $u_1$ can get it from either $u_2$ or $u_3$. To make our model simple, we always consider that the first *Standard* user getting the content, will always share it. In this example, $u_3$ shares $c$ among the other *Standard* users, not $u_2$. How the pricing policy changes when multiple users share the same content, i.e., when $u_2$ also shares $c$, is a part of future research direction. Also, nowadays, users have contracts with local network providers, in order to guarantee their connectivity to get internet access. That is why, in our model, we ignore the price a user has to pay to the NP just to download one media at one point of time. Furthermore, users also have contracts to one or more CP in order to get access to the content portfolio they offer (either their own content or from other CP as e.g., MGM), but in this paper we only focus on the interaction between a single CP and the users. A more complex model with multiple CPs and multiple users is another part of future research direction."

The parameters associated with this content distribution model are listed in Table I. In the most general case, we can assume that among all the users who download $c$, some users opt for the *Premium* option and some others for the *Standard* option, like the case we described in the previous example. In this example, only CP and $u_3$ own $c$ and can share it, so the bargaining will be only between CP and $u_3$. In that case, $Ben_{CP} = f.p_b + 2.f.p'_b - 2.f.p_n - f.p_u$ and $Ben_{user} = Ben_{u_3} = f.p_u - f.p'_b - f.s$. From the expressions of $Ben_{CP}$ and $Ben_{u_3}$ we observe that CP gets more benefit when more users download the content , and $u_3$ gets more benefit when he/she shares the content with more number of *Standard* users. The main goal of both the CP and $u_3$ is to maximize their benefits, i.e., finding $\max Ben_{CP}(\cdot)$ and $\max Ben_{user}(\cdot)$ respectively. Based on the values of $x$ and $y$, three cases can arise which are discussed in the next Section.

## IV. Proposed Pricing Policy

There can be three possible scenarios based on the value of $y$:

- **scenario 1**: $y = 0$, all of the $x$ users select *Standard* option. One user gets the content from the server and the others (i.e., $x - 1$) get the content from the first user.
- **scenario 2**: $y = x$, i.e., all of the $x$ users select the *Premium* option.
- **scenario 3**: $0 < y < x$, (as shown in Fig. 2 for the particular case of $x$=7 and $y$=3). This is the most general case.

In this Section, we consider each scenario separately and maximize $Ben_{CP_i}(\cdot)$ and $Ben_{user_i}(\cdot)$, where $Ben_{CP_i}(\cdot)$ and $Ben_{user_i}(\cdot)$ denote the benefit of CP and user respectively in the $i$-th scenario with $i \in \{1, 2, 3\}$. While doing so, we also calculate the values of $p_b$, $p'_b$ and $p_u$ in each scenario and check if equilibrium (a settlement of price to which both CP and user agree) can be obtained in each scenario.

### A. Case 1 : $y = 0$

In this case, all of the $x$ users select the *Standard* option. This option promotes content sharing and hence, we assume that one user gets the content from the server, and the remaining $x - 1$ get the content directly from him/her.

*1) CP:* In this case, the CP has to deliver the content to at least one user from whom all the remaining $x - 1$ users can share. The revenue of the CP is $f.p'_b.x$. The cost to deliver the content to one user is $f.p_n$. Furthermore, the CP has to give a reward to the user who is sharing the content with other interested users and thus helping in the distribution of the content. The amount of the reward is $f.p_u$ per sharing. So, the total amount of reward CP has to give is is $f.p_u.(x - 1)$ as the content is shared $(x - 1)$ times. So, in this case the total benefit generated by CP is :

$$Ben_{CP_1} = f.p'_b.x - f.p_n - f.p_u.(x - 1) \tag{2}$$

*2) User:* When a user downloads the content, it has to pay a price of $f.p'_b$ to the CP. The user also has to pay for the energy consumed while sharing the content with $x - 1$ users, which is given by $f.s.(x - 1)$. On the other hand, the user gets the reward from the CP, which is $f.p_u.(x - 1)$ for sharing the content $(x - 1)$ times.

So, the total benefit of the user can be expressed by

$$Ben_{user_1} = f.p_u.(x - 1) - f.p'_b - f.s.(x - 1) \tag{3}$$

It is important to note that selecting the *Standard* option while getting a content, gives to the user a chance to have positive benefit. This should motivate the user to choose this option over *Premium* option, which always generates a negative benefit.

*3) Nash Bargaining Solution (NBS):* Here we propose a bargaining between CP and users in a situation where:

- the value of $x \geq 2$,
- there is a conflict of interest on agreement, and
- there is a possibility of concluding a mutually beneficial agreement.

Considering $f$ and $p_n$ to be constant in this particular game, we can see from Eq. 2 that $Ben_{CP_1}$ is maximized when the function $F_{CP_1} = (p'_b.x - p_u.(x - 1))$ is maximized. From Eq. 3, we also see that $Ben_{user_1}$ is maximized when the following function $F_{user1} = (p_u.(x - 1) - p'_b)$ is maximized. To maximize $Ben_{CP_1}$, if the value of $p'_b$ is increased and the value of $p_u$ is decreased, value of $Ben_{user_1}$ decreases. On the other hand, to maximize $Ben_{user_1}$, if the value of $p_u$ is increased and the value of $p'_b$ is decreased, the value of $Ben_{CP_1}$ decreases. This is a classic bargaining problem and by using Nash Bargaining Solution (NBS), each of them can settle to a value which lies exactly in the middle of their interests. So, in this case, to attain equilibrium, both CP and user should settle to a value of

$$F_{eq_1} = \left( \frac{F_{CP1} + F_{user1}}{2} \right) = \frac{p'_b(x-1)}{2}$$

instead of $F_{CP_1}$ and $F_{user_1}$ respectively. Thus, at equilibrium, $F_{CP_1} = F_{user_1} = F_{eq_1}$. It is to noted that the value of $F_{eq_1}$ depends on $p'_b$ only and it is independent of $p_u$.

So, substituting the new values of $F_{CP_1}$ and $F_{user_1}$ to Eq. 2 and Eq. 3 respectively, the benefits of CP and user become

$$Ben_{CP_1} = f.\left( \frac{p'_b(x-1)}{2} - p_n \right) \quad (4)$$

$$Ben_{user_1} = f.\left( \frac{p'_b(x-1)}{2} - s(x-1) \right) \quad (5)$$

Now, $Ben_{CP_1}$ and $Ben_{user_1}$ are in equilibrium. By increasing the value of the same parameter ($p'_b$), both $Ben_{CP_1}$ and $Ben_{user_1}$ can be increased and there is no conflict of interest anymore.

*4) Settlement of Price:* We have already seen from Eq. 1 that $p'_b > p_n$. We also know that $p_n > 0$ and the value of $p_n$ is determined by the network provider. The upper limit on $p'_b$ can be set by studying the market and based on the customer's willingness to pay for the content. The benefits for both CP and users are now at equilibrium, as both $Ben_{CP_1}$ and $Ben_{user_1}$ are maximized when $F_{eq_1}$ is maximized.

*B. Case 2 : $y = x$*

In this case, all the $x$ users select the *Premium* option.

*1) CP:* The CP has to pay a total amount of $f.p_n.x$ to the NP to deliver the content to all $x$ users. In return, the CP earns a total amount of $f.p_b.x$ as revenue. So, in this case, benefit of CP is: $Ben_{CP_2} = f.(p_b - p_n).x$.

*2) User:* In this scenario, user pays a price of $f.p_b$ and does not generate any revenue, hence, the benefit of user in this scenario is: $Ben_{user_2} = -f.p_b$.

*3) Nash Bargaining Solution (NBS):* $Ben_{CP_2}$ is maximized when $p_b$ is maximized (as $f$, $p_n$ and $x$ are constants). On the other hand, $Ben_{user_2}$ is maximized when $p_b$ is minimum. As this is again a bargaining situation as described in Section IV-A3, equilibrium can be obtained when $p_b = 0$, which is very unrealistic.

*4) Settlement of Price:* As no realistic equilibrium (settlement of price) is not possible, CP selects the value of $p_b$ and the upper limit on this value can be set based on the customer's willingness to pay for the content. From the perspective of CP, to have a positive benefit, $p_b$ should be set at greater than $p_n$.

*C. Case 3 : $x > y > 0$*

This is the most general case, where from a group of $x$ users, $y$ users select the *Premium* option and the remaining $x-y$ select the *Standard* option. Here, as $y > 0$, so the lowest possible value of $y$ is $y = 1$ and as $x > y$, the lowest possible value of $x$ is $x = 2$.

*1) CP:* The total costs of CP in this case is: $f.p_n.y + f.p_n + f.p_u.(x-y-1)$. CP generates a revenue of $f.p_b.y + f.p'_b(x-y)$. benefit of CP in this case is:

$$Ben_{CP_3} = f.p_b.y + f.p'_b.(x-y) - f.p_u.(x-y-1) - f.p_n.(y+1)$$

*2) User:* The cost associated to the user selecting the *Standard* option and sharing the content with the remaining $x-y-1$ users is: $f.p'_b + s.f.(x-y-1)$. On the other hand, the revenue of such a user is: $f.p_u.(x-y-1)$. The user's benefits in this case can be expressed by:

$$Ben_{user_3} = f.p_u.(x-y-1) - f.p'_b - s.f.(x-y-1)$$

*3) NBS:* In this case, $Ben_{CP_3}$ is maximized when $F_{CP_3} = p_b.y + p'_b.(x-y) - p_u.(x-y-1)$ is maximized, whereas $Ben_{user_3}$ is maximized when $F_{user_3} = p_u.(x-y-1) - p'_b$ is maximized. As there is again a conflict of interest between $F_{CP_3}$ and $F_{user_3}$, using NBS equilibrium can be obtained the same way as described in Section IV-A3. Proceeding the same way as described in Section IV-A3, the equilibrium benefits of CP and user are:

$$Ben_{CP_3} = f\left( \frac{p_b.y + p'_b.(x-y-1)}{2} - p_n.(y+1) \right) \quad (6)$$

$$Ben_{user_3} = f.\left( \frac{p_b.y + p'_b.(x-y-1)}{2} - s.(x-y-1) \right) \quad (7)$$

*4) Settlement of Price:* The benefits of both CP and users are maximized when the term $F_{eq_3} = f.\left( \frac{p_b.y + p'_b.(x-y-1)}{2} \right)$ is maximum. We use Lagrange multiplier [16] to obtain the maximum value for $F_{eq_3}$ with the constraint $p_b > p'_b$. Let us assume $p_b - p'_b = \delta$ where $\delta \in \mathbb{R}^+$. We introduce a new variable called Lagrange multiplier ($\lambda$) and study the Lagrange function defined by

$$\mathsf{L} = f.(p_b.y + p'_b.(x-y-1)) + \lambda.(p_b - p'_b - \delta) \quad (8)$$

The maximum is obtained by setting the derivatives (Eq. 8) to zero.

$$\frac{\partial \mathsf{L}}{\partial p_b} = f.y + \lambda = 0 \quad (9)$$

and

$$\frac{\partial \mathsf{L}}{\partial p'_b} = f.(x-y-1) - \lambda = 0 \quad (10)$$

Eq. 9 leads to $\lambda_{max} = -f.y$, and putting this value of $\lambda_{max}$ in Eq. 10, we get $f.(x-y-1) = -f.y$, i.e., $x = 1$, which contradicts the basic condition as we already stated that lowest possible value of $x$ in this case is 2. This proves that we cannot have a maximum value for $F_{eq_3}$ with the constraints $x \geq 2$, $y \geq 1$, $x > y$ and $p_b > p'_b$.

Let us discuss how $Ben_{CP_3}$ varies in a simple example. Let us define $r$ as the ratio between *Premium* and *Standard* service costs, i.e., $r = \frac{p_b}{p'_b} = \frac{m}{n}$. Obviously, $r > 1$. Let us define as $k$, the percentage of users getting the *Premium* service, i.e., $y = k.x$. In this case, $Ben_{CP_3}$ can be expressed as

$$Ben_{CP_3} = \frac{f.p_n}{2}\left( n.(r.k.x - x - k.x - 1) - (k.x+1) \right)$$

Let us evaluate the dependence of $Ben_{CP_3}$ with respect $r$, $k$ and $x$. We use cost units (c.u.) as the basic unit of the $p_n$ cost when transmitting $f = 1$ Byte. In all the simulations we assume $x = 10$ and $n = 2$ unless stated otherwise.

Fig. 4a shows the variation of $Ben_{CP_3}$ with respect $k$ and $r$. It can be observed that for small $k$, i.e., small number of *Premium* users, $r$ does not influence the benefits. However, as the percentage of *Premium* users increases, the impact of $r$

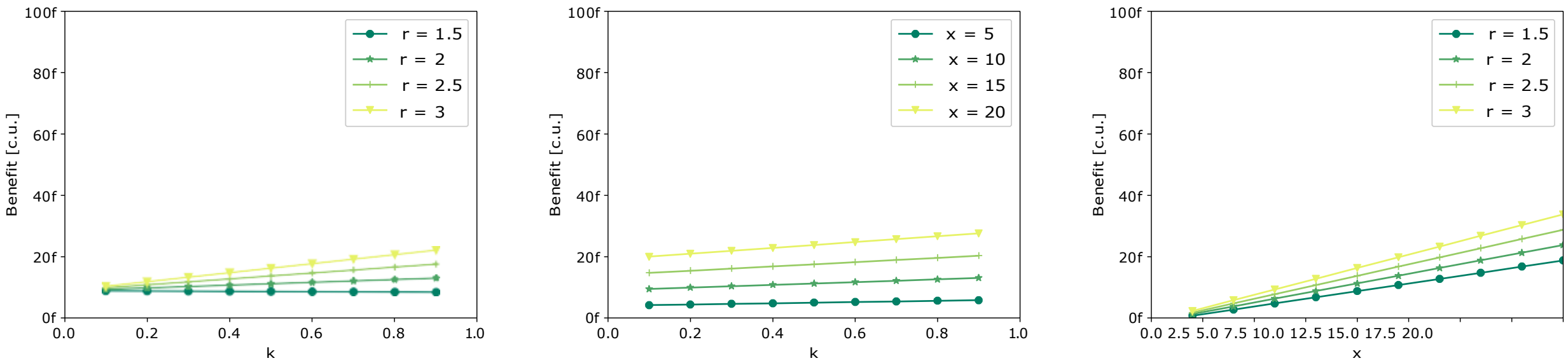

*(a)* Benefit vs $k$ for different $r$ *(b)* Benefit vs $k$ for different $x$ *(c)* Benefit vs $x$ for different $r$

Fig. 4: $Ben_{CP\,3}$ dependence for varying parameters in a simple scenario.

also increases (the higher $r$, the higher the increase). Fig. 4c evaluates $Ben_{CP_3}$ with respect $k$ and $x$ for $r = 2$. The figure shows that $x$ plays a major role since even for low $k$ values, the benefits increases significantly with $x$ and they do not vary substantively with $k$. Lastly, Fig. 4b depicts the benefits for different $r$ and $x$ values, keeping $k = 0.5$. It can be observed that small $x$ limits the benefits independently of $r$. However, for large numbers of users, the benefits increases and more significantly, for high $r$ values.

### *D. Comparison among $Ben_{CP_1}$, $Ben_{CP_2}$ & $Ben_{CP_3}$*

When a user gets the content directly from server, the load on the server increases and CP also has to take care of the content delivery, which includes a lot of other costs. On the other hand, when users share content among them, CP does not always have to take care of the content delivery and it costs CP much lesser. This is why we assume that $p_b > p'_b$.

From Eq. 1, we know that $p'_b > p_n$. Hence, we can state that $p_b > p'_b > p_n$. As the value of $p_n$ is beyond control of both CP and users, let us use $p_n$ as the smallest unit of money and express $p_b$ and $p'_b$ in terms of $p_n$:

$$p_b = m.p_n \qquad p'_b = n.p_n \tag{11}$$

where $m$ and $n$ are considered to be positive real numbers with $m > n$ (as we already assumed $p_b > p'_b$ at an earlier point) and $n > 1$ (as we already know from Eq. 1 that $p'_b > p_n$). Considering these values in the benefit of CP, given in Eq. 4, it can be expressed as

$$Ben_{CP_1} = f.p_n.\left(\frac{n(x-1)}{2} - 1\right)$$

Let us compare this benefit with the benefit that the CP would get in Case 2 ($Ben_{CP_2}$). Subtracting $Ben_{CP_1}$ from $Ben_{CP_2}$, we get

$$\begin{aligned} Ben_{CP_2} - Ben_{CP_1} &= f.p_n.(m-1).x - f.p_n\left(\frac{n.(x-1)}{2} - 1\right) \\ &= f.p_n\left(\frac{(m-n).x + (m-2).x + n + 2}{2}\right) \end{aligned} \tag{12}$$

Let us find the values of $m$ for which the value of Eq. 12 is greater than 0:

- As $m > n$, when $m \geq 2$, the value of Eq. 12 is greater than 0.
- When $m < 2$, the value of Eq. 12 is still greater than 0 if
$$\begin{aligned} (m-n) &> (2-m) \\ \text{or, } 2m &> (n+2) \end{aligned} \tag{13}$$
As $n > 1$, $2m > 3$ or $m > 1.5$. Thus, the value of Eq. 12 is greater than 0 when $m > 1.5$.
- When $m \leq 1.5$, $2m <= 3$. As we already saw in the previous step, the value of Eq. 12 is greater than 0 when $2m > (n + 2)$. Combining these two conditions we get $(n + 2) \leq 3$ or $n \leq 1$, which directly violates Eq. 1.

Thus, we can see that for $m > 1.5$, $Ben_{CP_2}$ is always greater than $Ben_{CP_1}$.

Similarly, subtracting $Ben_{CP_3}$ from $Ben_{CP_2}$, we get,

$$\begin{aligned} Ben_{CP_2} - Ben_{CP_3} &= f.p_n.(m-1).x \\ &- f.p_n.\left(\frac{m.y + n.(x-y-1)}{2} - p_n.(y+1)\right) \\ &= f.p_n\left(\frac{(2m-n-2).x + (n+2).y + n + 2}{2}\right) \end{aligned} \tag{14}$$

Thus, $Ben_{CP_2} > Ben_{CP_3}$ holds when $2m > (n + 2)$, which is same as Eq. 13.

From these calculations in this section, we can conclude that for $m > 1.5$, $Ben_{CP_2}$ is the highest among $Ben_{CP_1}$, $Ben_{CP_2}$ and $Ben_{CP_3}$.

## V. Real life study

To study our proposed solution in a real life scenario, we consider a peer-to-peer content distribution network formed by Wi-Fi signals [7]. We use a real life dataset of the BBC iPlayer, which is one of the most popular catch-up TV services in the UK, used by over 60% of the total population. We analyze the content access pattern using BBC iPlayer trace file of July 2014 which consists of around 215 million sessions from 25 million distinct users and 17 million different IP addresses [16]. We randomly select 100 users from them and study their content access history. Out of the 100 users we select, we assume $x$ among them use premium option and the rest of them use standard option, where $x$ can vary between 0 and 100. We randomly select $x$ number of users from the set of 100 users, do this random selection in a confidence interval of 95% and plot the benefit of the CP for various values of $x$. The benefit is expressed as $f.p_n.10^9$ where $f$ = 1GB.

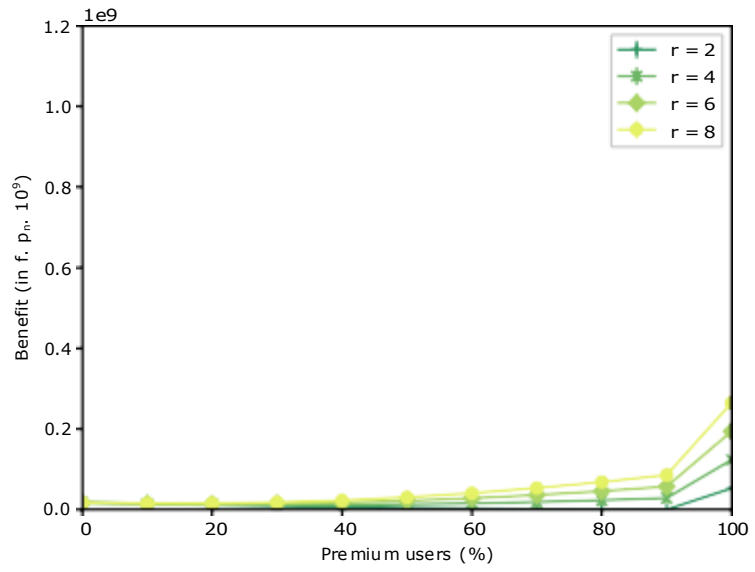

(a) $n = 2$

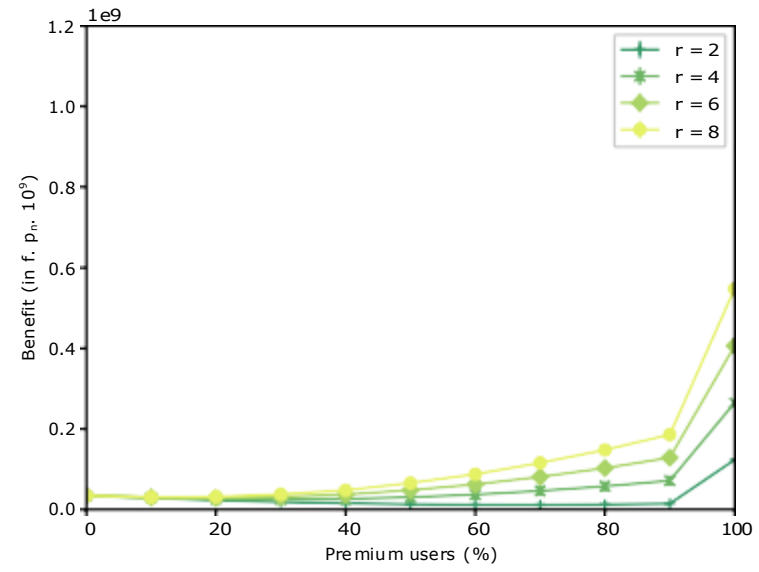

(b) $n = 4$

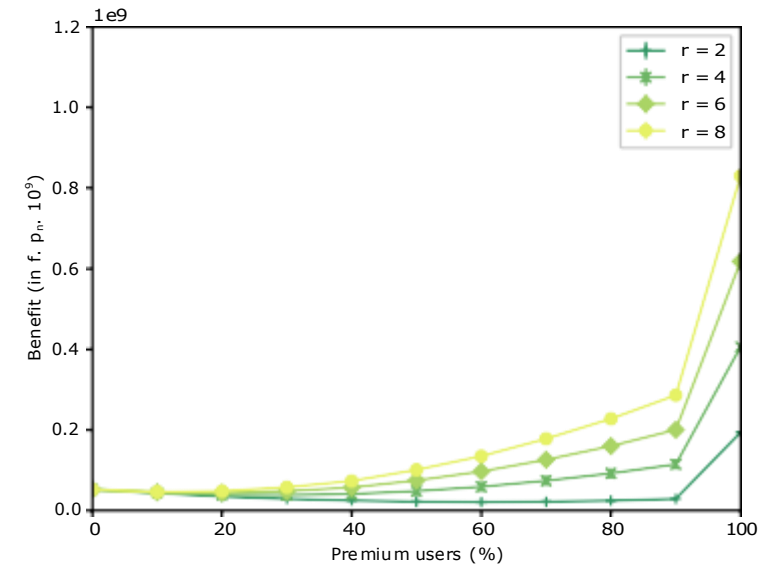

(c) $n = 6$

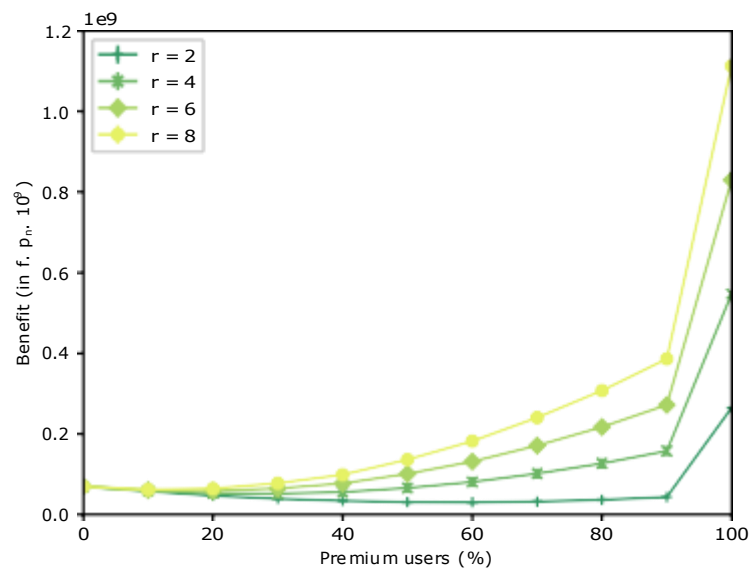

(d) $n = 8$

Fig. 5: $Ben$ dependence for varying parameters in a real life scenario.

From Fig. 5 we see that when the value of $n$ increases, the benefit of the CP also increases, because both $p_b$ and $p^J_b$ becomes much greater than $p_n$. Also, from all plots in Fig. 5 we see that when the value of $r$ is low, i.e., the value of $p_b$ and $p^J_b$ are within close range of each other, the benefit of the CP gradually decreases when the percentage of premium user increases because as the percentage of standard users is decreasing, content sharing among them is also reducing while the cost to put the content at one user remains the same, which in return reduces the overall benefit of the CP. But when the percentage of standard users becomes 0, there is no cost invloved for the CP to put the content, and that is why we notice a sharp increase in the benefit between $y = 90$ and $y = 100$.

From these plots it is obvious that the higher the value of $n$ gets, the more benefit CP gets. Also, if CP wants to make profit independent of the percentage of premium users, the value of $r$ should be greater than or equal to 4. We also observe that the benefit of the CP is maximum when the percentage of premium users is highest (100%). So, from the perspective of CP, the most profitable content distribution model is when he and only he has the control over his content and he can determine the price accordingly. But the problem with this centralized content distribution model has already been discussed in Section I. That is why in the next Section we try to trade off between these two model and obtain the revised pricing policy taking these factors into account.

## VI. Extension of Pricing Policy

The CP has two main important objectives to consider:

- maximize the benefits ($Ben_{CP}$), and
- minimize the load on the server ($LR$).

| Scenario | $Ben_{CP}$ | $LR$ |
|---|---|---|
| Scenario 1 | $f.\frac{p^J_b.(x-1)-2p_n}{2}$ | $\frac{x-1}{x}$ |
| Scenario 2 | $f.(p_b - p_n).x$ | 0 |
| Scenario 3 | $f.\frac{p_b.y+p^J_b.(x-y-1)-2p_n(y+1)}{2}$ | $\frac{x-y-1}{x}$ |

TABLE II: CP's benefits and load reduction for the three cases

The benefit of CP and the load reduction on the server for the three previous cases presented in Section IV, have been summarized in Table II.

CP has to prioritize one of these two objectives, because as we have already seen in Section IV-D, the benefit for CP is maximum when $m > 1.5$, and in that scenario, reduction of load on the server is minimum (i.e., $LR = 0\%$). The lesser the load on the server, the lesser is the maintenance cost. Let us then assume, for simplicity, a proportional model of the extra benefits with respect the load reduction: A load reduction of a factor of $g$, implies an increase of CP's benefit equal to $\varphi g$, where $\varphi > 0$ is the proportionality constant. At the end of this section we discuss how the value of $\varphi$ can be calculated.

Let us also assume that CP wants to balance between these two factors, and accordingly the CP gives weights $\alpha$ and $\beta$ to benefits and load respectively, such that $\alpha + \beta = 1$. In this way, the CP wants to maximize his utilization by maximizing the following target function $TF_{CP} = \alpha.Ben_{CP} + \beta.\varphi.LR$. The value of the weights - $\alpha$ and $\beta$, are to be determined by only CP. Let us consider three cases based on the value of $\alpha$ and $\beta$:

### A. Case 1 : $\alpha = 0, \beta = 1$

In this case, CP only aims at reducing the load on its server. According to Table II, the load reduction is maximized in Scenario 2 and minimized for Scenario 1.

### B. Case 2: $\alpha = 1, \beta = 0$

This case is considered when the CP wants to maximize the benefit. This case is the one discussed in detail in Section IV, along with the $p_b$ and $p'_b$ values for the three different scenarios.

### C. Case 3: $\alpha > 0, \beta > 0$ and $\alpha + \beta = 1$

In Section IV we discussed the values of $p_b$ and $p'_b$ in three different cases. However, when CP launches a new media content, CP does not know in advance the values of $x$ and $y$. So, CP should have such a policy such that the value of its $TF_{CP}$ always remains the same and the value if independent of the values of $x$ and $y$. It happens only if the value of the $TF_{CP}$ is same in all of the three cases.

If we equalize the $TF_{CP}$ of Case 1 and Case 2, we get:

$$a.f.p_b.x = a.f.p'_b.\frac{(x-1)}{2} + (a.p_n x.f - a.p_n.f + \beta.\varphi.\frac{(x-1)}{x}) \quad (15)$$

Similarly equalizing the $TF_{CP}$ of Case 2 to Case 3 and combining it with Eq. 15, we obtain

$$p_b = p'_b + 2p_n + \frac{2\beta\varphi}{afx} \quad (16)$$

As $p_n > 0, \beta > 0, \alpha > 0$ and $f > 0$, $p_b > p'_b$, which is the basic motivation of our solution to pursue more users to choose the standard option. If we consider that the large number of users get the content, i.e., large value of $x$, we can approximate $\frac{1}{x} \to 0$. Thus, $p_b = p'_b + 2.p_n$, and considering Eq. 11, it implies

$$m = n + 2 \quad (17)$$

Since $m = r.n$, Eq. 17 leads to $n.(r-1) = 2$. In this case, $n$ is maximum when $(r-1)$ is minimum. As we already know $r > 1$, let us assume that $r = 1 + \delta$, where $\delta > 0$, $\delta_{max} \to \infty$ and $\delta_{min} \to 0$. As $(r-1)_{min} = \delta_{min}$, we get $n_{max} = \frac{2}{\delta_{min}}$ or $n_{max} \to \infty$, and $m_{max} = 2 + \frac{2}{\delta_{min}}$ or $m_{max} \to \infty$. Similarly, as $(r-1)_{max} = \delta_{max}$, we get $n_{min} = \frac{2}{\delta_{max}}$ or $n_{max} \to 0$, and $m_{min} = 2 + \frac{2}{\delta_{max}}$ or $m_{max} \to 2$. This gives us bounds on the values of $m$ and $n$.

On the other hand, putting the value of $p_b$ from Eq. 16 into Eq. 15 we get

$$a.f.(p'_b + 2p_n).x = a.f.p'_b.\frac{(x-1)}{2} + a.p_n.f(x-1) + \beta.\varphi.\frac{(x-1)}{x}$$

As the value of $x$ is quite high, we can write $x - 1 \approx x$ and $x + 1 \approx x$. After rearranging the terms, we find

$$p'_b + 2p_n = \frac{1}{2}p'_b + p_n + \frac{\beta\varphi}{af}$$
$$\text{or, } p'_b = -2p_n + 2\frac{\beta\varphi}{af} \quad (18)$$

Putting this value of $p'_b$ in Eq. 16 we get $p_b = 2\frac{\beta\varphi}{af}$. Hence, the values of $p_b$ and $p'_b$ are independent of $x$. Also,

$$\frac{1}{r} = \frac{p'_b}{p_b} = 1 - \frac{p_n af}{\beta\varphi}$$

Based on the value of $\frac{p'_b}{p_b}$, once the values of $f$, $p_n$, $\alpha$ and $\beta$ are known, the value of $\varphi$ can also be easily calculated.

## VII. Conclusion

We considered a digital media delivery system consisting of two main players: user and Content Provider (CP). CP launches the content in his digital platform so that users get it either directly from the server or from another user who has already downloaded it. Based on the number of users using each option, we considered three different cases and determine the price of the content in each case. We used Nash Bargaining Solution (NBS) to determine the prices and discussed if equilibrium can be obtained in each case. We considered CP to have mainly two objectives: maximizing the benefit and reducing the load on the server. The load on the server can be reduced if users share content among themselves. In this paper we provided a pricing policy from the perspective of CP in three different cases based on the priority of CP.

There are a few directions in which this work can be further extended. In the peer-to-peer network model, we considered only one user would cache the content and would share with others. As a next step, multiple content caching and sharing scenario can be considered. Also, we considered in the *standard* option, a user always gets the content from another user. It can be also studied how the pricing varies when the user does not always get the content from another user, instead he has to get it directly from the server.